# DETECTORS FOR THE FUTURE OF X-RAY IMAGING

M. Åslund [a],[*], E. Fredenberg [b], M. Telman [a], M. Danielsson [b]
[a]Sectra Mamea AB, Smidesvägen 2, 171 41 Solna, Sweden, [b]Department of Physics, Royal Institute of Technology, AlbaNova, 106 91, Stockholm, Sweden

In recent decades, developments in detectors for x-ray imaging have improved dose efficiency. This has been accomplished with e.g. structured scintillators such as columnar CsI, or with direct detectors where the x-rays are converted to electric charge carriers in a semiconductor. Scattered radiation remains a major noise source, and fairly inefficient anti-scatter grids are still a gold standard. Hence, any future development should include improved scatter rejection. In recent years, photon-counting detectors have generated significant interest by several companies as well as academic research groups. This method eliminates electronic noise, which is an advantage in low-dose applications. Moreover, energy-sensitive photon-counting detectors allow for further improvements by optimizing the signal-to-quantum-noise ratio, anatomical background subtraction, or quantitative analysis of object constituents. This paper reviews state-of-the-art photon-counting detectors, scatter control and their application in diagnostic x-ray medical imaging. In particular, we will focus on spectral imaging with photon-counting detectors, evaluate such pitfalls as charge sharing and high rates, and discuss various proposals for mitigation.

## INTRODUCTION

The goal of this paper is to review state-of-the-art photon-counting detectors, scatter control and their application in diagnostic x-ray medical imaging. In recent decades, there has been great improvement in physical image quality and dose efficiency in radiography. This has been accomplished through improved efficiency in the detectors and scatter rejection mechanisms. The most recent technology to be introduced was photon-counting in a scanning multi-slit geometry. Even with improved physical image quality of the projection image, diagnostic accuracy can be limited by superimposed tissue. This limitation can be overcome with the use of computed tomography (CT), tomosynthesis, or spectral imaging. In these applications, a significant amount of research and development has recently been performed by several companies and academic research groups.

## PHOTON-COUNTING DETECTORS

The chronology of digital technologies introduced in radiography starts with indirect computed radiography (CR) systems nearly three decades ago, followed by indirect flat-panels using CsI as scintillator and direct a-Se flat-panels.[1] The a-Si flat-panel has a photodiode array that is used to record the light from a CsI scintillator deposited on top of it. The CsI can be manufactured with a columnar structure to guide the light towards the photodiode beneath. Instead of a scintillator, direct detectors use an a-Se photoconductor deposited on the flat-panel, and charges are thereby created directly at the interaction site of absorbed x-ray photons. An a-Si thin-film transistor array with storage capacitors is used to accumulate the charge at each pixel. The spread of charges is less than the spread of light, and higher resolution can be expected in direct detectors. In multi-row CT, dedicated indirect detectors are mainly used. Cone-beam CT using indirect flat-panel detectors is gaining interest for such applications as bone and lung imaging, mammography, and kV imaging for radiation therapy.[2]

In mammography and CT, the latest advancement is to use direct photon-counting detectors. Such detectors have been used in medical imaging for decades in PET and SPECT applications. The introduction in x-ray imaging has been delayed, mainly due to higher count-rates and stricter requirements for detector resolution. The result is that mammography, with its relatively low count-rates per pixel, was the first application with a photon-counting system to be commercialized, in 2003.[3] Photon-counting detectors for multi-row CT are being developed in academic and industrial research groups.[4–7] Si[3,4,8] and CdTe / CdZnTe[5–7] dominate as photo-conductors, although other detector materials are used as well, e.g. noble gases.[9]

Photon-counting electronics use a threshold to discriminate charge pulses from the noise floor of detector and electronics, i.e. photon-counting is implemented as pulse-counting. One of the major benefits of counting each photon compared with accumulating the total electrical charge is that the detector noise can be completely rejected in the detection. A second advantage of photon-counting technology is that, since pulse height is proportional to photon energy, the detector can be made energy-sensitive if additional thresholds are introduced.

---

[*]magnus.aslund@sectra.se

The challenge presented by the high count-rates is pile-up, which means that the detector is too slow to distinguish between two or more consecutive photons. These are then counted as a single photon with energy equal to or higher than that of the individual photons, resulting in reduced detector efficiency and energy resolution. A first-order approximation for the reduction in efficiency is $1 - count\text{-}rate \times dead\text{-}time$. Readout electronics used in state-of-the-art photon-counting systems have dead-times of approximately 200 ns which, with the typical count-rates in photon-counting mammography of less than 50 MHz/mm$^2$ and typical pixel size of 50 $\mu$m results in an efficiency loss of less than 2.5%.[10] In CT, the count-rates are in the range 100-1000 MHz/mm$^2$ and the pixels are substantially larger than in mammography. If these count-rates are not accounted for, neither the electronics nor the detector material would be fast enough to avoid significant losses due to pile-up. Mitigating strategies include increasing the speed of the electronics and reading out smaller detector elements than required in terms of resolution.

When the detector elements are reduced in size, charges from one photon interaction may be divided between two adjacent photon-counters' electrodes. This is referred to as charge sharing and is a major challenge of photon-counting. Charge sharing correlates the image noise to some extent and reduces spatial resolution because a fraction of the photons are double-counted. The energy resolution in spectral imaging is also affected because two photons of lower energy are recorded instead of the original one photon. Mitigating strategies include avoiding small pixels and using anti-coincidence logic in the readout-electronics. Anti-coincidence can be implemented in a relatively straightforward manner in one-dimensional detectors that are read out individually,[10] whereas logic to avoid ambiguities for multi-hit events in two-dimensional detectors requires readout electronics faster than state-of-the-art.[11]

CdTe and CdZnTe have high effective atomic numbers and consequently a high absorption efficiency. The process of making crystals of these materials is complicated, however, and the detectors are generally expensive and suffer from imperfections in the crystal structure. The latter leads to charge trapping, resulting in pile-up due to slow charge diffusion and reduced energy resolution since all charge does not reach the electrodes. In addition, the high atomic number leads to problems with fluorescence if the pixels are smaller than a few hundred microns. Fluorescent photons that escape the pixel reduce the energy resolution and create an escape peak in the detected energy spectrum. If fluorescent photons are detected in adjacent pixels, double-counting occurs, resulting in problems similar to those encountered with charge sharing. Crystalline Si detectors are commonly used in many fields of physical research. The material has efficient charge diffusion, and fluorescence is generally negligible. Its low atomic number, however, causes low absorption efficiency, and the absorption length of Si in the medical imaging range is in the order of one mm. There is also a relatively large amount of Compton scattering in Si. Both of these problems are most prominent at high photon energies such as in CT.[4,10] The low absorption efficiency is generally addressed by placing Si-strip detectors in an edge-on geometry.[11]

The detective quantum efficiency (DQE) is the efficiency figure of merit for digital detectors. The peak DQE has evolved from 35% or lower with screen-film systems to 45% or higher with the digital flat-panel systems and 68% with the photon-counting system.[12] With an increased DQE, the physical image quality is improved at maintained radiation dose or the radiation dose can be decreased at maintained physical image quality. The connection between physical image quality and diagnostic accuracy is not straightforward, in part because the imaging task includes anatomical structures. Efforts aimed at developing low noise detectors with high sensitivity are motivated by radiation dose considerations in applications ranging from conventional radiography to spectral CT, where in the latter applications the detector must be efficient not only because of the low dose used per projection in tomography, but also because the dose is further divided into the separate energy bins.

## SYSTEM PERFORMANCE

The DQE can be defined as the noise equivalent quanta (NEQ) over the detected number of quanta had an ideal (photon-counting) detector been used, and accordingly it does not account for any loss of large-area contrast. The signal-difference-to-noise ratio (SDNR) is the figure of merit for physical image quality in digital x-ray imaging.[13] In addition to the NEQ, the SDNR includes the subject contrast $C$ and is given by $\text{SDNR}^2 = C^2 \text{NEQ}$. The figure of merit used for the efficiency of digital systems when the system's contrast transfer is also included is often referred to as dose efficiency (DE). It is defined as the squared SDNR achieved with the system ($\text{SDNR}_{\text{out}}^2$) over the squared SDNR had an ideal system been used ($\text{SDNR}_{\text{in}}^2$), i.e.[14]

$$\text{DE}(\rho) = \frac{\text{SDNR}_{\text{out}}^2(\rho)}{\text{SDNR}_{\text{in}}^2} = \frac{C_{\text{out}}^2}{C_{\text{in}}^2} \frac{\text{NEQ}(\rho)}{Q_{\text{in}}}, \qquad (1)$$

were $\rho$ is the spatial frequency. Thus, the DE is determined by the combination of the large-area contrast transfer function and the signal-to-quantum-noise ratio transfer function.

Usually, the noise in $\text{SDNR}_{\text{out}}$ refers to the sum of quantum noise and any additive noise, where the latter is often close to zero for a photon-counting system. For many imaging tasks, however, quantum noise is not the limiting factor for detectability, because lesions are obscured by anatomical structures with a frequency spectrum approximately equal to an inverse power function. The anatomical structures can be approximated as random anatomical noise and included in a generalized $\text{SDNR}_{\text{out}}$ to find a generalized DE (GDE) via Eq. (1).[15]

It is important also to consider how observer and object weigh different spatial frequencies when interpreting the $\text{DE}(\rho)$ or $\text{GDE}(\rho)$.[13,15] For instance, large objects are relatively affected by the anatomical noise, whereas at higher frequencies the quantum noise dominates. Several observer models exist, which take observer and object into account to find a task dependent dose efficiency. In particular, the ideal observer is the integral of Eq. (1) with a signal template, and represents the upper limit of observer performance.[13] It should also be noted that depending on the specific task, a large part of the anatomical structure may be deterministic, and therefore does not disturb the observer.

## SCATTERED RADIATION

Scattered radiation is a major source of image quality degradation in x-ray imaging, and should be compared to the efforts put into detector performance. The secondary radiation leads to a decrease in the subject contrast. It also has a negative impact on reconstructions and quantitative analysis. Methods to reduce the level of scattered radiation include air gaps, grids and scanning techniques.[16] The grids are still widely used today even after the transition to digital detectors. To quantify the level of scattered radiation, the scatter-to-primary ratio (SPR) is used, where "primary" refers to radiation that has not been scattered in the object. In Table 1, the SPR for some typical applications is shown, and for many applications the SPR is greater than 1, which means that the contrast is reduced by nearly half. With the SDNR, we can relate the loss of contrast to the loss in NEQ in the detector, e.g. $\geq 50\%$ for state-of-the-art detectors. The scatter device efficiency (SDE) is the analogue to the detector DQE and is defined as the square of the SDNR with the device to the square of the SDNR assuming ideal scatter rejection. It then follows that $\text{SDE} = \text{SDF} \cdot T_p$, where $T_p$ is the primary transmission and $\text{SDF} = 1/(1 + \text{SPR})$ is the scatter degradation factor.[17] $\text{SDF}_0$ is the degradation factor when no means of scatter rejection is used and without scatter rejection, $\text{SDE} = \text{SDF}_0$. In Table 1, the $\text{SDF}_0$ for some typical applications and the SDE of a typical grid are shown. The grid is the "medium" grid from Neitzel et al[16] with $T_p = 0.7$ and selectivity $\Sigma = 6$. The following observations can be made from the table: (1) the loss of image quality due to inefficiencies in the grid is comparable to or worse than that of state-of-the-art detectors. (2) With the most efficient scatter rejection techniques, an SDE close to 1 can be achieved, e.g. scanning multi-slit.[17] Thus, the SDE and therefore the system's dose efficiency can be increased by 40% in mammography or even by a factor of four in a pelvic radiograph.

Table 1. Scatter-to-primary ratio (SPR) adapted from Aichinger et al,[18] scatter degradation factor without a grid ($\text{SDF}_0$) and scatter device efficiency for a "medium" grid from Neitzel et al[16] ($\text{SDE}_{\text{grid}}$) for various examinations.

| Examination | SPR | $\text{SDF}_0$ | $\text{SDE}_{\text{grid}}$ |
|---|---|---|---|
| Mammography | 0.50 | 0.67 | 0.65 |
| Skull | 0.82 | 0.55 | 0.62 |
| Lung | 1.22 | 0.45 | 0.58 |
| Cardiac | 2.3 | 0.30 | 0.50 |
| Pelvis | 4.0 | 0.20 | 0.42 |

## SPECTRAL IMAGING

Spectral imaging refers to imaging of the x-ray energy dimension, sometimes referred to as the "x-ray color". X-ray attenuation is material-specific because of different dependence on the atomic number for the photoelectric and Compton cross sections, and discontinuities in the photoelectric cross sections at absorption edges. Consequently, spectral imaging can extract information about the object constituents.[19,20] Scattered radiation is particularly problematic for spectral imaging since, apart from reducing the contrast, it also reduces the energy resolution, and efficient rejection schemes are thus necessary.

There are at least three potential benefits of spectral imaging compared to non energy-resolved imaging:

1. Energy weighting refers to optimization of SDNR and thereby DE with respect to its energy dependence; photons at energies with larger target-to-background contrast can be assigned a greater weight.[7,21,22] Energy-integrating detectors have intrinsic energy weighting proportional to the energy, and photon-counting detectors without energy resolution weigh all photons equally.

2. Background subtraction or dual-energy subtraction, is approximately equal to optimization of the GDE. Because x-ray attenuation is material specific, a weighted subtraction of two images acquired at different mean energies cancels the contrast between any two materials, whereas all other materials remain visible to some degree.[8,23,24] The contrast in the subtracted image is greatly improved if the lesion is enhanced by a contrast agent with an absorption edge in the energy interval.

3. Material decomposition refers to the extraction of information about the object, e.g. differentiation, quantification, etc.[6,19,20,25] This option can to some extent be regarded a generalization of the two above.

Several solutions to obtain spectral information are being pursued; switching of beam quality (kVp and filter),[23,26] sandwich detectors,[27,28] and two different-beam-quality x-ray sources with two corresponding detectors.[29,30] Results are promising, but the effectiveness of these approaches may be impaired due to overlap of the spectra, a limited flexibility in choice of spectra, additional scatter if two sources illuminate the object simultaneously, increased risk of motion artifacts when switching between spectra, and a limited number of energy levels (in practice only two).

A solution to the mentioned challenges may be to instead use photon-counting detectors, which potentially provide higher energy resolution, a larger number of energy levels, and no need for several exposures or sources.[5–8,24] Because only one x-ray source is required, there is also a cost advantage, although a two-source solution for a photon-counting detector to speed up the data collection is also conceivable.

Table 2. Technology, peak DQE, scatter device efficiency (SDE), spectral efficiency (SE) and dose efficiency (DE) for a microcalcification. The DE is normalized to one for the photon-counting system.

| Technology[a] | DQE | SDE | SE | DE |
|---|---|---|---|---|
| Indirect CR (Mo/Rh) | 0.45 | 0.62 | 0.62 | 0.40 |
| Indirect CsI (Mo/Rh) | 0.42 | 0.62 | 0.62 | 0.46 |
| Direct Se (Mo/Rh) | 0.40 | 0.70 | 0.62 | 0.50 |
| Direct Se (W/Ag) | 0.40 | 0.70 | 0.75 | 0.60 |
| Photon-counting (W/Al) | 0.68 | 0.96 | 0.64 | 1.00 |

[a]Values for W/Ag assumes 75 μm Ag filter, Mo/Rh assumes 25 μm Rh and W/Al assumes 0.5 mm Al. SE of W/Rh with 50 μm Rh is within 3% of the W/Ag combination. The indirect systems' SDE is for a linear grid, the direct systems SDE is a cellular grid and the photon-counting system's SDE is for a multi-slit geometry.

## APPLICATIONS

### Radiography

In all applications but mammography, photon-counting remains in the research and development phase. In this section, a calculation of the DE in mammography of photon-counting and state-of-the-art scatter control relative other technologies will therefore be summarized. The DQE in mammography by Monnin et al[12] includes the DQE of an indirect CR system (Fuji CR Profect), an indirect flat-panel using CsI as scintillator (GE Senographe DS), a direct a-Se flat-panel (Lorad Selenia) and a photon-counting system (Sectra MicroDose D40). In mammography, the SDE has been published for the linear and cellular grids used by the flat-panel detectors and for the scanning multi-slit geometry used by the photon-counting system.[17] In Table 2, the peak DQE[12] (at 50 μGy) is tabulated for different detector technologies, as is the SDE[17] for the applicable scatter rejection technology for a breast thickness of 5 cm. For completeness, the table includes applicable factors for spectral efficiency (SE).[31] The SE factors include the effect of energy weighting efficiency.[22] In Fig. 1, the resulting DE as defined in Eq. 1 is shown calculated by multiplying[14,32] the frequency dependent DQE[12] and the non-frequency dependent SDE and SE factors from Table 2.

In Table 2, the dose efficiency for an ideal observer, calculated with a bowl-shaped object of diameter 0.24 mm resembling a microcalcification, is shown. This bowl diameter corresponds to the fourth speck group in the ACR accreditation phantom. These results show that with the photon-counting system a dose reduction in the 40-60% range is possible with maintained physical image quality.

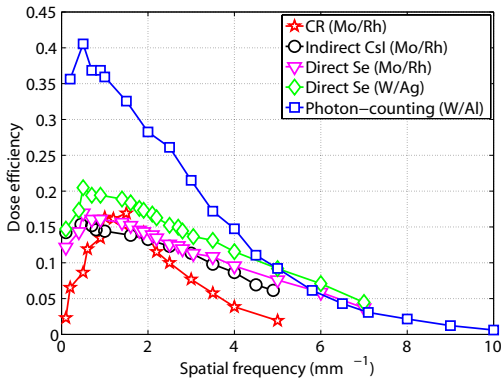

Figure 1. Dose efficiency of indirect CR, indirect CsI, direct Se and photon-counting.

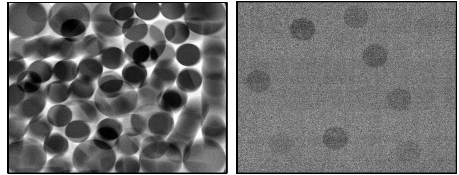

Figure 2. Images of a phantom with iodine and anatomical clutter, acquired using a photon-counting mammography system with an energy-sensitive detector.[33] **Left:** Absorption image. **Right:** Dual-energy subtracted image.

In spectral mammography, most research efforts to date have been directed towards energy weighting and background subtraction. For imaging of soft tissue in mammography, optimal energy weighting is approximately the inverse cube of the photon energy and it can improve the DE by approximately 30% compared with energy integrating detectors, and 20% compared with non energy-resolved photon-counting detectors.[22]

Dual-energy subtraction mammography with photon-counting detectors has been investigated and found feasible for enhancement of tumors.[8] Contrast enhancement with iodine might increase the contrast in the subtracted image if the spectrum covers the K-edge at 33.2 keV. Phantom measurements, with an example shown in Fig. 2, suggest that dual-energy subtraction may increase an ideal observer dose efficiency for tumor detection $\sim$ 60% compared to absorption imaging.[33] The minimum visible iodine concentration in the cited study was $\sim$ 3 mg/ml for 3 mm cavities (0.9 mg/cm$^2$), where 3-4 mg/ml is an expected uptake for breast tumors. Simulations indicate that the improvement for dense breasts may be as high as a factor of six, and relatively large improvements are foreseen for an optimized detector. If perfect background subtraction is assumed, the standard DE can be used as a figure of merit. In that case, a photon-counting detector with perfect energy resolution shows an improvement compared to a dual-spectra approach of 145%.[34]

## Computed Tomography

Three-dimensional imaging, e.g. CT, can reduce superimposed anatomical noise, and hence improve the GDE, however, at the cost of increased radiation dose to achieve an adequately low quantum noise level. Scattered radiation is problematic for three-dimensional reconstruction, and leads to localized artifacts such as streaking and cupping artifacts, and inaccuracies in quantitative analysis, e.g. reconstructed CT numbers.[2] Therefore, mechanical methods to reduce the scattered radiation that reaches the detector (e.g. grids) are necessary, and post-processing of the data is often used as a complement.

Photon-counting has the advantage of high low-dose performance, which is necessary to reduce the patient dose in CT, and there has been a significant increase in photon-counting CT efforts over the last few years;[4–7] the first clinical images were presented only in 2008 at the RSNA meeting, and promising results for pre-clinical applications have been shown. CdZnTe is most often used as a detector material, but Si in an edge-on geometry may be an interesting alternative for less pile-up and better scatter rejection.[4]

Optimal energy weighting in spectral CT has been shown to improve the DE for calcifications and iodine by 40% and 60%, respectively, compared to energy integrating techniques.[7] A wider and perhaps even more promising application of spectral imaging, however, is material decomposition, i.e. differentiation and quantification of object constituents. Promising results have been achieved in terms of e.g. discrimination between and classification of contrast agents, plaque, and kidney stones, reduction of artifacts from heavy substances (stents, amalgam etc.), and reduction of beam-hardening artifacts.[6,7,35] Materials with an absorption edge in the spectral range of CT ($\sim$30 to 90 keV) can be uniquely identified. The K-edge of iodine at 33.2 keV is on the low end of the spectrum, and cannot be used except perhaps for lungs, small children, and thin patients in general. Gadolinium on the other hand has a conveniently positioned K-edge at 50 keV and may be the preferred contrast agent for some applications in photon-counting CT. It remains, however, to prove that the required concentrations are nontoxic.

Limited angle tomography, i.e. tomosynthesis, is a compromise between dose and depth resolution. The modality has been the focus for much research since the introduction of flat-panel detectors in the 1990s. Photon-counting is being evaluated for breast tomosynthesis with Si[36] and gaseous detectors.[9] Although overlapping tissue is not a problem in CT, minimization of the background clutter contrast is believed to be beneficial in tomosynthesis.

CONCLUSION

Photon-counting, and energy-discriminating photon-counting is pursued in radiography and CT. The detectors can be made efficient due to rejection of detector noise when counting individual photons. Efficiency losses due to scattered radiation are comparable to losses in state-of-the-art detectors, so any future development should consider the effects of scattered radiation. Efficient scatter rejection and photon-counting detectors result in dose-efficient systems, which allow for low-dose imaging, e.g. 40-60% of the dose relative other technologies in mammography. Energy sensitive photon-counting detectors allow for further improvements by optimizing signal-to-quantum-noise ratio, anatomical background subtraction, or quantitative analysis of object constituents.